\documentclass[preprint,12pt]{elsarticle}

\usepackage{graphicx}

\usepackage{amssymb}

\begin{document}

\begin{frontmatter}



\title{f(R) Gravity and Maxwell Equations from the Holographic Principle}

\author[label2]{Rong-Xin Miao}
\author[label1]{Jun Meng}
\author[label1]{Miao Li}
\address[label1]{Kavli Institute for Theoretical Physics, Key Laboratory
of Frontiers in Theoretical Physics, Institute of Theoretical
Physics, Chinese Academy of Sciences, Beijing 100190, People's
Republic of China}
\address[label2]{Interdisciplinary Center for Theoretical Study,
University of Science and Technology of China, Hefei, Anhui 230026,
People's Republic of China}

\begin{abstract}
Extending the holographic program of \cite{Li}, we derive f(R) gravity and the
Maxwell equations from the holographic principle, using time-like holographic screens. We find that to
derive the Einstein equations and f(R) gravity in a natural
holographic approach, the quasi-static condition is necessary. We
also find the surface stress tensor and the surface electric
current, surface magnetic current on a holographic screen for f(R)
gravity and Maxwell's theory, respectively.

\end{abstract}


\end{frontmatter}


\section{Introduction}

Gravity may not be a fundamental force, this may be related to a deep principle underlying
the working of our world: The holographic principle. The most recent concrete proposal
for a macroscopic picture of gravity is due to Verlinde \cite{Verlinde} (see also \cite{Pad}), which in turn
is motivated by the gauge/gravity correspondence in string theory and an earlier proposal
of Jacobson \cite{Jacobson}.

However, there are problems with the Verlinde's proposal, as pointed out in \cite{Li}.
First, Verlinde's derivation of the Einstein equations always requires a positive
temperature on a holographic screen, this is not attainable for an arbitrary screen. To avoid
this problem, we propose to use a screen stress tensor to replace the assumption of equal partition of
energy. Second, using Verlinde's proposal, we do not have a reasonable holographic thermodynamics,
there is always an area term in the holographic entropy of a gravitational system. There is no
such problem in our proposal. A prediction of the program of \cite{Li} is that there is always
a huge holographic entropy associated to a gravitational system, with a form similar to the
Bekenstein bound.

In this paper, we would like to clarify the role of the adiabatic condition used in \cite{Li} left
unexplained in that paper. To see how far our program can get, we also use the same idea
to derive f(R) gravity, we see that unlike Verlinde's proposal in which it is
impossible to accommodate theories containing higher derivatives \cite{lipang}, there is a natural ansatz for the surface
stress tensor to derive the f(R) theory.

Lastly, using the same idea with a surface current replacing the surface stress tensor, we derive the
Maxwell equations. Our success demonstrates that our program is more universal.

This paper is organized as follows. In sect.2, we review our
holographic derivation of the Einstein equations on a time-like
screen. In sect.3 and sect.4, we derive the f(R) equations and the
Maxwell equations in a similar holographic approach, respectively.
We conclude in sect.5.

\section{Einstein equations from the holographic principle}

In this section, we shall review our derivation of the Einstein
equations in a holographic program \cite {Li} different from
Verlinde's \cite{Verlinde}. We will explain the physical reason to
impose the quasi-static condition used in the previous paper
\cite{Li} in order to derive the Einstein equations naturally on
a time-like holographic screen.

Let us start with some definitions. Our holographic screen is a 2+1
dimensional time-like hyper-surface $\Sigma$, which can be open or
closed, embedded in the 3+1 dimensional space-time $M$.  We use $x^a$,
$g_{ab}$, $\nabla_{a}$, $y_i$, $\gamma_{ij}$ and $D_{i}$ (here
$a$, $b$ run from 0 to 3, and $i$, $j$ run from 0 to 2) to denote the
coordinates, metric and covariant derivatives on $M$ and $\Sigma$,
respectively.
\begin{figure}
\includegraphics[scale=0.36]{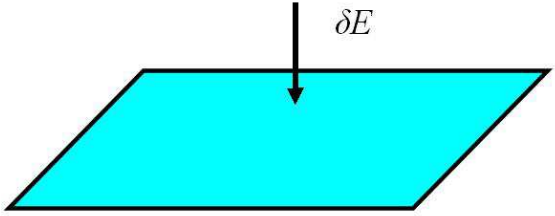}
\caption{An energy flux $\delta E$ passing through an open patch on
the holographic screen. } \label{fig1}
\end{figure}

Unlike Jacobson's idea \cite{Jacobson}, we consider an energy
flux $\delta E$ passing through an open patch on a {\it time-like} holographic
screen $d\Sigma=dAdt$ (see Fig. \ref{fig1})
\begin{equation}\label{E1}
 \delta E=\int_{\Sigma} T_{ab}\xi^{a}N^{b}dA dt,
\end{equation}
where $T_{ab}$ is the stress tensor of matter in the bulk $M$,
$\xi^a$ is a time-like Killing vector, and $N^a$ is the unit
vector normal to $\Sigma$. To define $N^a$ we may assume $\Sigma$ be
specified by a function
\begin{equation}\label{fSigma}
f_{\Sigma}(x^{a})=c .
\end{equation}
Thus, the normalized normal vector to $\Sigma$ is
\begin{equation}\label{Na}
N^{a}=\frac{g^{ab}\nabla_{b}f_{\Sigma}}{\sqrt{\nabla_{b}f_{\Sigma}\nabla^{b}f_{\Sigma}}}
.
\end{equation}

As in \cite{Li}, we introduce the surface energy density $\sigma$ and the surface energy
flux $j$ on the screen. These are quantities contained in the surface stress
tensor $\tau^{ij}$ and given by
\begin{equation}\label{sigma}
 \sigma=u_{i}\tau^{ij}\xi_{j}\ ,\ j=-m_{i}\tau^{ij}\xi_{j},
\end{equation}
where $u_{i}$, $m_{i}$ are the unit vectors normal to the screen's
boundary $\partial\Sigma$ (we will give the expressions of $u_i$ and
$m_i$ shortly), $\xi_{i}$ is a Killing vector on
$\Sigma$. In a quasi-static space-time $u^{i}$ is related to $\xi^{i}$
by $u^{i}=e^{-\phi}\xi^{i}$, and $\phi$ is the Newton's potential
defined by $\phi=\frac{1}{2}\log(-\xi^i\xi_i)$ on the screen. Note
that $\sigma$ and $j$ are the energy density and energy flux on the
screen measured by the observer at infinity. Apparently, central to
our discussion is the choice of $\tau_{ij}$. Naturally, $\tau^{ij}$
should depend on the extrinsic geometry of the screen, since
extrinsic geometry contains both the information of the bulk $M$ and of
the screen $\Sigma$, thus it is a natural bridge relating both sides
to realize the holographic principle.
 Thus we assume the following simplest and most general form
\begin{equation}\label{t}
 \tau^{ij}=n K^{ij}+  q \gamma^{ij},
\end{equation}
where $n$ is a constant, $q$ is a function to be determined and $K^{ij}$ is the
extrinsic curvature on $\Sigma$ defined by
\begin{equation}
 K_{ij}=-e^{a}_{i}e^{b}_{j}\nabla_{a}N_{b},
\end{equation}
where $e^{b}_{j}=\frac{\partial x^{a}}{\partial
 y^{i}}$ is the projection operator satisfying $N_{a}e^{a}_{i}=0$.

On the screen, the change of energy has two sources, one is due to
variation of the energy density $\sigma$, the other is due to energy
flowing out the patch to other parts of the screen, given by the
energy flux $j$. The energy variation on the patch is then given by
\begin{eqnarray}\label{EE2}
\delta E&=&\int(u_{i}\tau^{ij}\xi_{j})dA|^{t+dt}_{t}-\int
m_{i}\tau^{ij}\xi_{j}dydt\nonumber \\
&=&-\int_{\partial\Sigma}(M_{i})\tau^{ij}\xi_{j}\sqrt{h}dz^{2}=-\int_{\Sigma}D_{i}(\tau^{ij}\xi_{j})dAdt
,
\end{eqnarray}
\begin{figure}
\includegraphics[scale=0.36]{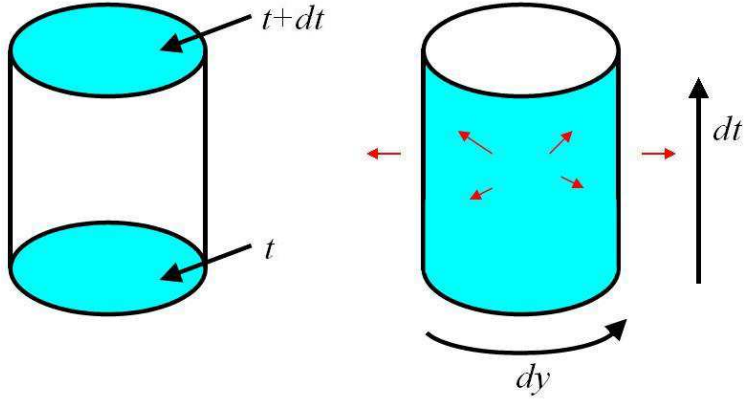}
\caption{Left panel: The first term
$\int(u_{i}\tau^{ij}\xi_{j})dA|^{t+dt}_{t}$ in Eq.(\ref{EE2}) is due
to change of the density. Right panel: The second term $-\int
m_{i}\tau^{ij}\xi_{j}dydt$ is the energy flow (denoted by the red
arrows) through the patch boundary parameterized by $y$. }
\label{fig2}
\end{figure}
where the first term in the first equality is due to change of the
density and the second term is the energy flow through the patch
boundary parameterized by $y$ (see Fig. \ref{fig2}). These two terms
can be naturally written in a uniform form, since the boundary of
the patch consists of two space-like surfaces ($d\Sigma$ at $t$ and
$t+dt$), and a time-like boundary. $h$ is the determinant of the
reduced metric on $\partial\Sigma$, $D_{i}$ is the covariant
derivative on $\Sigma$. $M_{i}$ is a unit vector in $\Sigma$ and is
normal to $\partial\Sigma$. Let us choose a suitable function
$f_{\partial\Sigma}(y^{i})=c$ on
 $\Sigma$ to denote the boundary $\partial\Sigma$, then we have
\begin{equation}\label{Mi}
M^{i}=\frac{\gamma^{ij}D_{j}f_{\partial\Sigma}}{\sqrt{D_{j}f_{\partial\Sigma}D^{j}f_{\partial\Sigma}}}.
\end{equation}
Notice that when $M^{i}$ is along the direction of $dy^{0}$($dt$) it
becomes $u^{i}$, and when along the direction of
$dy^{i}$($dy^{1},dy^{2}$) it becomes $m^{i}$.

For a reason to be clear later, we focus on a quasi-static process in the following
derivations. Recall that in Eq.(\ref{fSigma}) we use
$f_{\Sigma}(x^{a})=c$ to denote the holographic screen $\Sigma$. In
the quasi-static limit, $f_{\Sigma}(x^{a})$ is independent of time,
so
$N_{a}\sim(0,\partial_{1}f_{\Sigma},\partial_{2}f_{\Sigma},\partial_{3}f_{\Sigma})$.
Note that $\xi^{a}\simeq(1,0,0,0)$ in the quasi-static limit, thus
we have $N_{a}\xi^{a}\rightarrow 0$. The Killing vector $\xi_{i}$ on
$\Sigma$ can be induced from the Killing vector $\xi_a$ in the bulk
$M$ in the quasi-static limit:
\begin{equation}
 \xi_{i}=\xi_{a}e^{a}_{i},\ \
 D_{(i}\xi_{j)}=\nabla_{a}(\xi_{b}-N_{c}\xi^{c}N_{b})e^{a}_{(i}e^{b}_{j)}=K_{ij}N_{a}\xi^{a}\rightarrow
 0 .
\end{equation}
From Eq.(\ref{t}) and the Gauss-Codazzi equation
$R_{ab}N^{a}e^{b}_{i}=-D_{j}(K^{j}\ _i-K\gamma^{j}\ _i)$, one can
rewrite Eq.(\ref{EE2}) as
\begin{eqnarray}\label{E2}
\delta E&=&\int_{\Sigma}[nR_{ab}\xi^{i}e^{a}_{i}N^{b}-\xi^{i}D_i(n
K+q)]dAdt\nonumber \\
&=&\int_{\Sigma}[nR_{ab}\xi^{a}N^{b}-\xi^{a}\nabla_a(n K+q)]dAdt,
\end{eqnarray}
where we have used the formulas $\xi^a=\xi^{i}e^{a}_{i}$,
$\xi^{i}D_i f=\xi^{a}\nabla_a f$ in the quasi-static limit. It
should be stress that the second term $\xi^{a}\nabla_a(n K+q)$ in
Eq.(\ref{E2}) can not be written in the form $\xi^{a}N^{b}B_{ab}$
($B_{ab}$ is independent of $\xi^{a}$ and $N^{b}$). For details,
please refer to the Appendix.

Take into consideration that $\xi^a$, $N^b$ are independent vectors
and $g_{ab}\xi^{a}N^{b}=0$, equating Eq.(\ref{E1}) and Eq.(\ref{E2}) results in
\begin{equation}\label{Einstein}
nR_{ab}+fg_{ab}=T_{ab},\ \ \ q=-nK,
\end{equation}
where $f$ is an arbitrary function. We note that the second equation is a consequence of the fact
that the second term on the R.H.S. of Eq.(\ref{E2}) must be vanishing. This result tells us
that the surface stress tensor is the same as the Brown-York surface stress tensor, but we have
not made use of any action. Using the energy conservation
equation $\nabla^aT_{ab}=0$, we obtain $f=n(-\frac{R}{2}+\Lambda)$.
Defining the Newton's constant as $G=\frac{1}{8\pi n}$, we then get
the Einstein equations
\begin{equation}\label{Einstein1}
R_{ab}-\frac{R}{2}g_{ab}+\Lambda g_{ab}=8\pi G T_{ab}.
\end{equation}
Substituting $n=\frac{1}{8\pi G}$ and $q=-n K$ into Eq.(\ref{t}), we
obtain
\begin{equation}\label{Brownt}
\tau^{ij}=\frac{1}{8\pi G}(K^{ij}- K\gamma^{ij}),
\end{equation}
this is just the quasi-local stress tensor of Brown-York defined in
\cite{Brown}. Now we have derived the Einstein equations and the
Brown-York stress tensor from our holographic program.

In the above derivations, for simplicity, we have imposed the
quasi-static condition $N_a\xi^a=0$. We now briefly discuss the
relation between the quasi-static condition and the holographic
principle. Let us try to see what will happen when we abandon the
quasi-static condition $N_a\xi^a=0$. The energy flux $\delta
E$ passing through an open patch on the holographic screen
$d\Sigma=dAdt$ becomes
\begin{equation}\label{EEE1}
 \delta E=\int_{\Sigma} T_{ab}\xi^{a}N^{b}dA dt=\int_{\Sigma} [T_{ab}e^{a}_i\xi^iN^{b}+T_{ab}N^aN^b(N_c\xi^c)]dA dt,
\end{equation}
a new term $T_{ab}N^aN^b(N_c\xi^c)$ which is proportional to the
pressure along the direction of $N^a$ appears. Since now we aim to
study the relationship between the quasi-static condition and the
holographic principle, not to derive the Einstein equations, for
simplicity, we assume that the Einstein equations are satisfied and
so the surface stress tensor $\tau_{ij}$ is the Brown-York stress
tensor Eq.(\ref{Brownt}). Then, Eq.(\ref{EEE1}) is expected to be
\begin{eqnarray}\label{Energy1}
 \delta E&=&\int_{\Sigma}\frac{1}{8\pi G} [R_{ab}e^{a}_i\xi^iN^{b}+(R_{ab}-\frac{R}{2}g_{ab})N^aN^b(N_c\xi^c)]dA
 dt\nonumber \\
 &=&\int_{\Sigma} [-\xi_jD_i\tau^{ij}+\frac{1}{16\pi G}(- ^{(3)}R+K^2-K_{ij}K^{ij})(N_c\xi^c)]dA
 dt.
\end{eqnarray}
We have used the Gauss-Godazzi equations
\begin{eqnarray}\label{Gauss}
 & &R_{ab}N^{a}e^{b}_{i}=-D_{j}(K^{j}\ _i-K\delta^{j}\ _i), \nonumber \\
 & &(2R_{ab}-Rg_{ab})N^aN^b=- ^{(3)}R+K^2-K_{ij}K^{ij},
\end{eqnarray}
where $^{(3)}R$ is the Ricci scalar on $\Sigma$.

Note that $\xi_i=e^a_i\xi_a$ is on longer a Killing vector on the
screen, since $D_{(i}\xi_{j)}=K_{ij}N_{a}\xi^{a}\neq0$. If we still
assume Eq.(\ref{sigma}) with $\xi_i$ being the projection of $\xi^a$
on $\Sigma$, the change of energy on screen Eq.(\ref{EE2}) becomes
\begin{eqnarray}\label{Energy2}
\delta E&=&\int(u_{i}\tau^{ij}\xi_{j})dA|^{t+dt}_{t}-\int
m_{i}\tau^{ij}\xi_{j}dydt\nonumber \\
&=&-\int_{\partial\Sigma}(M_{i})\tau^{ij}\xi_{j}\sqrt{h}dz^{2}=-\int_{\Sigma}D_{i}(\tau^{ij}\xi_{j})dAdt\nonumber \\
&=&\int_{\Sigma}[-\xi_jD_i\tau^{ij}-\tau^{ij}D_i\xi_j]dAdt \nonumber \\
&=&\int_{\Sigma}[-\xi_jD_i\tau^{ij}+\frac{1}{8\pi G}(K^2-K^{ij}K_{ij})(N_c\xi^c)]dAdt \nonumber \\
\end{eqnarray}
Since $-^{(3)}R-K^2+K^{ij}K_{ij}\neq0$, the second term of
Eq.(\ref{Energy1}) is not equal to the second term of
Eq.(\ref{Energy2}) except when $N_c\xi^c=0$. Of course, one can add
some terms to Eq.(\ref{Energy2}) to equate Eq.(\ref{Energy1}) and
Eq.(\ref{Energy2}), but it is very unnatural. So if we expect the
holographic principle to require that the the energy flow through
the holographic screen (defined by Eq.(\ref{Energy1})) and the
change of energy on the holographic screen (defined by
Eq.(\ref{Energy2})) are equal to each other, the quasi-static
condition $(N_a\xi^a\rightarrow0)$ is necessary.

Finally, even without using the Einstein equation, the fact that the second term
in  Eq.(\ref{EEE1}) is proportional to the transverse component of $T_{ab}$ tells us that
we need to introduce an transeverse term in the surface stress tensor, this is not
holographic at all.

\section{f(R) gravity from the holographic principle}

 In this section, we shall derive the equations of f(R) gravity in the same manner as in
 \cite{Li} and  the previous section. For the same reason as in the presiouc section we impose the quasi-static condition
$(N_a\xi^a\rightarrow0)$.

 The energy flux $\delta E$ passing through an open patch on the holographic
 screen $d\Sigma=dAdt$ and the change of energy on the screen take the same form  as in Eq.(\ref{E1}) and Eq.(\ref{EE2})
\begin{equation}\label{fE1}
 \delta E=\int_{\Sigma} T_{ab}\xi^{a}N^{b}dA dt,
\end{equation}
\begin{eqnarray}\label{fEE2}
\delta E&=&\int(u_{i}\tau^{ij}\xi_{j})dA|^{t+dt}_{t}-\int
m_{i}\tau^{ij}\xi_{j}dydt\nonumber \\
&=&-\int_{\partial\Sigma}(M_{i})\tau^{ij}\xi_{j}\sqrt{h}dz^{2}=-\int_{\Sigma}D_{i}(\tau^{ij}\xi_{j})dAdt.
\end{eqnarray}
The only difference is the assumption of surface stress tensor
$\tau^{ij}$. As argued in Sect.2, $\tau^{ij}$ should depend on the
the extrinsic geometry of the screen, since
it is a natural bridge relating the bulk $M$ and the holographic screen $\Sigma$.
Instead of the simplest assumption Eq.(\ref{t}),
we now assume more generally
\begin{equation}\label{ft}
 \tau^{ij}=n f'(R)K^{ij}+  q \gamma^{ij},
\end{equation}
where $f'(R)$ is a general function of the Ricci scalar $R$, and $q$ is to be determined. The case $f'(R)=1$ is special and is what
underlying the Einstein equations.
Substituting Eq.(\ref{ft}) into Eq.(\ref{fEE2}), we obtain
\begin{eqnarray}\label{fE2}
\delta E&=&-\int_{\Sigma}D_{i}(\tau^{ij}\xi_{j})dAdt \nonumber \\
&=&\int_{\Sigma}(nf'R_{ab}N^{a}\xi^{b}-n\xi^{a}K_{ab}\nabla^{b}f'-\xi^{a}\nabla_{a}(q+nf'K)) dAdt \nonumber\\
&=&\int_{\Sigma}(nf'R_{ab}N^{a}\xi^{b}+n\xi^{a}\nabla_{a}N_{b}\nabla^{b}f'-\xi^{a}\nabla_{a}(q+nf'K)) d A dt\nonumber\\
&=&\int_{\Sigma}[nN^{a}\xi^{b}(R_{ab}f'-\nabla_{a}\nabla_{b}f')+\xi^{a}\nabla_{a}(nN^{b}\nabla_{b}f'-nf'K-q)]d A dt.\nonumber\\
\end{eqnarray}
In the above calculation, we have used the Gauss-Codazzi equation $R_{ab}N^{a}e^{b}_{i}=-D_{j}(K^{j}\ _i-K\gamma^{j}\ _i)$, $\xi^a\nabla_aR=0$,
$\xi^{a}K_{ab}=-\xi^{a}\nabla_{a}N_{b}$ and the quasi-static condition $(N_a\xi^a\rightarrow0)$.
As the case in Sect.2 the second term of Eq.(\ref{fE2}) $\xi^{a}\nabla_{a}(nN^{b}\nabla_{b}f'-nf'K-q)$ does not contain term $\xi^{a}N^{b}B_{ab}$,
where $B_{ab}$ is independent of $\xi^{a}$ and $N^{b}$. For details, please refer to the Appendix (just replace $(n K+q)$ by $(nf'K+q-nN^{b}\nabla_{b}f')$.

Since $\xi^a$, $N^b$ are independent vectors
and $g_{ab}\xi^{a}N^{b}=0$, equating Eq.(\ref{fE1}) and Eq.(\ref{fE2}) results in
\begin{eqnarray}\label{}
& &q=nN^{b}\nabla_{b}f'-nf'K,\ \nonumber\\
& &f'R_{ab}-\nabla_a\nabla_bf'+Fg_{ab}=\frac{1}{n}T_{ab},
\end{eqnarray}
where $F$ is an arbitrary function, using $\nabla_aT^{ab}=0$, we get
\begin{eqnarray}\label{F}
\nabla^bF=\nabla^b(\nabla_c\nabla^cf'-\frac{f}{2}).
\end{eqnarray}
Thus, $F=\Box f'-\frac{f}{2}+\Lambda$. Define the Newton's constant as $G=\frac{1}{8\pi n}$, we then obtain
equations of motion of f(R) gravity
\begin{eqnarray}\label{f(R)}
f'R_{ab}-\frac{f}{2}g_{ab}-\nabla_a\nabla_bf'+g_{ab}\Box f'+\Lambda g_{ab}=8\pi G T_{ab}.
\end{eqnarray}
Substituting $q=\frac{1}{8\pi G}(N^{b}\nabla_{b}f'-f'K)$ into Eq.(\ref{ft}), we
get the surface stress tensor of f(R) gravity
\begin{equation}\label{ft1}
 \tau^{ij}=\frac{1}{8\pi G} [f'(R)(K^{ij}-K\gamma^{ij})+ N^c\nabla_c f' \gamma^{ij}].
\end{equation}

Now, we have obtained the f(R) equations and surface stress tensor from our holographic program. Let us continue to understand the
physical meaning of the surface stress tensor Eq.(\ref{ft1}). It is well known that the action of f(R) gravity
\begin{equation}\label{Sf}
 S=\frac{1}{2\kappa}\int d^4x\sqrt{-g}f(R)+S_M(g_{ab},\psi)
\end{equation}
is equivalent to the Einstein gravity with a scalar field
\begin{equation}\label{CSf}
S=\int d^4x\sqrt{-\tilde{g}}[\frac{\tilde{R}}{2\kappa}-\frac{1}{2}\partial^a\tilde{\phi}\partial_a\tilde{\phi}-V(\tilde{\phi})]+
S_M(e^{-\sqrt{2\kappa/3}}\tilde{g_{ab}},\psi),
\end{equation}
if we perform conformal transformation $\tilde{g}_{ab}=f'g_{ab}$ and $\tilde{\phi}=\sqrt{\frac{3}{2\kappa}}\ln f'$, $V(\tilde{\phi})=\frac{Rf'-f}{2\kappa(f')^2}$.
The surface stress tensor for metric $\tilde{g_{ab}}$ is
\begin{equation}\label{Ct}
\tilde{\tau}^{ij}=\frac{1}{\kappa}(\tilde{K}^{ij}-\tilde{K}\tilde{\gamma}^{ij}).
\end{equation}
Note that $\tilde{N}_a=(f')^{1/2}N_a$, $\tilde{N}^a=(f')^{-1/2}N^a$, $\tilde{\gamma}_{ab}=f'\gamma_{ab}$ and $\tilde{\Gamma}^d_{cb}=\Gamma^d_{cb}+C^d_{cb}$,
where $C^{d}_{cb}=\frac{1}{2f'}(\delta^d_c\nabla_{b}f'+\delta^d_b\nabla_{c}f'-g_{cb}g^{de}\nabla_{e}f')$. After some calculation, we find
\begin{eqnarray}\label{Ct1}
\tilde{\tau}_{ij}&=&\frac{(f')^{1/2}}{\kappa}(K_{ij}-K\gamma_{ij})+\frac{(f')^{-1/2}}{\kappa} \gamma_{ij}N^d\nabla_df'\nonumber\\
&=&(f')^{-1/2}\tau_{ij}.
\end{eqnarray}
It is interesting that the surface stress tensors $\tilde{\tau}_{ij}$ and $\tau_{ij}$ of Eq.(\ref{ft1}) are related exactly by an conformal factor $(f')^{-1/2}$.
Let us move on to understand this conformal factor $(f')^{-1/2}$. Firstly, let us derive some useful formulas.
\begin{eqnarray}\label{formula}
&&\tilde{T}^M_{ab}=-2\frac{1}{\sqrt{-\tilde{g}}}\frac{\delta S_M}{\delta \tilde{g}^{ab}}=\frac{1}{f'}T^M_{ab}\nonumber\\
&&\tilde{T}^{\tilde{\phi}}_{ab}=\tilde{\nabla}_{a}\tilde{\phi}\tilde{\nabla}_{b}\tilde{\phi}
-\frac{1}{2}\tilde{g}_{ab}\tilde{g}^{cd}\tilde{\nabla}_{c}\tilde{\phi}\tilde{\nabla}_{d}\tilde{\phi}-\tilde{g}_{ab}V(\tilde{\phi})\nonumber\\
&&\tilde{N}_a=(f')^{1/2}N_a, \ \tilde{\gamma}_{ab}=\tilde{g}_{ab}-\tilde{N}_a\tilde{N}_b=f'\gamma_{ab}\nonumber\\
&&\tilde{M}_i=(f')^{1/2}M_i, \ \tilde{h}_{ij}=\tilde{\gamma}_{ij}-\tilde{M}_i\tilde{M}_j=f' h_{ij} \nonumber\\
&&\tilde{\xi}^a=\xi^a
\end{eqnarray}
We shall prove the last equation $\tilde{\xi}^a=\xi^a$ below. Assume
$\xi^a$ be a Killing vector of the metric $g_{ab}$, we have
\begin{equation}\label{}
L_{\xi}g_{ab}=\xi^c\partial_cg_{ab}+(\partial_a\xi^c)g_{cb}+(\partial_b\xi^c)g_{ca}=0,
\end{equation}
then
\begin{eqnarray}\label{}
L_{\xi}\tilde{g}_{ab}&=&\xi^c\partial_c(f'g_{ab})+f'(\partial_a\xi^c)g_{cb}+f'(\partial_b\xi^c)g_{ca}\nonumber\\
&=&g_{ab}\xi^c\partial_cf'+f'(\xi^c\partial_cg_{ab}+(\partial_a\xi^c)g_{cb}+(\partial_b\xi^c)g_{ca})\nonumber\\
&=&0,
\end{eqnarray}
where we have used the identity $\xi^a\nabla_a R=0$. Now, it is clear that $\xi^a$ is also a Killing vector of $\tilde{g}_{ab}$.
The energy flux $\delta E$ passing through the holographic
screen is
\begin{eqnarray}\label{fE21}
 \delta \tilde{E}&=&\int_{\Sigma} (\tilde{T}^M_{ab}+\tilde{T}^{\tilde{\phi}}_{ab})\xi^{a}\tilde{N}^{b}\sqrt{-\tilde{\gamma}}d^2xdt\nonumber\\
 &=&\int_{\Sigma} \tilde{T}^M_{ab}\xi^{a}\tilde{N}^{b}\sqrt{-\tilde{\gamma}}d^2xdt\nonumber\\
 &=&\int_{\Sigma} T_{ab}\xi^{a}N^{b}\sqrt{-\gamma}d^2x dt\nonumber\\
 &=&\delta E.
\end{eqnarray}
Above we have used again the identity $\xi^a\nabla_a R=0$ and the quasi-static limit
$(N_a\xi^a\rightarrow0)$, so that $\tilde{T}^{\tilde{\phi}}_{ab}\xi^{a}\tilde{N}^{b}=0$.

The change of energy on the screen is
\begin{eqnarray}\label{fE22}
\delta \tilde{E}&=&-\int_{\partial\Sigma}\tilde{M}^i\tilde{\tau}_{ij}\xi^{j}\sqrt{\tilde{h}}d^2z=-\int_{\partial\Sigma}M^i(f')^{1/2}\tilde{\tau}_{ij}\xi^{j}\sqrt{h}d^2z
\end{eqnarray}
Equating Eq.(\ref{fE21}) and Eq.(\ref{fE22}), we find
\begin{eqnarray}\label{}
\delta E&=&-\int_{\partial\Sigma}M^i(f')^{1/2}\tilde{\tau}_{ij}\xi^{j}\sqrt{h}d^2z\nonumber\\
&=&-\int_{\partial\Sigma}M^i\tau_{ij}\xi^{j}\sqrt{h}d^2z
\end{eqnarray}
It is clear that $(f')^{1/2}\tilde{\tau}_{ij}$ plays the rule of $\tau_{ij}$.

The above discussion is a check that from the holographic principle we have derived the correct surface stress tensor (\ref{ft1}) of $f(R)$ gravity,
and helps us to gain some insight into the physical meaning of f(R) surface stress tensor (\ref{ft1}): It is related with it's
conformal counterpart by an appropriate conformal factor $\tau_{ij}=(f')^{1/2}\tilde{\tau}_{ij}$.

\section{Maxwell equations from the holographic principle}

 In this section, we shall derive the Maxwell equations in a similar holographic
 approach as in the above two sections. However, there are two main differences. First, instead of using  conservation of energy,
 we use conservation of charge to derive the Maxwell equations. We calculate the charge passing through the holographic
screen in the bulk $M$ and the charge change on the screen $\Sigma$
respectively, and equate them as dictated by the holographic
principle. With an appropriate assumption of the current on the
screen, we can obtain the Maxwell equations. Secondly, we do not
need the quasi-static condition $(N_a\xi^a\rightarrow0)$, in fact we
make no use of a Killing vector in our derivations of the Maxwell
equations, since a Killing vector is related to  energy instead of
charge.

 Consider the electric charge $\delta Q$ passing through an open patch on the holographic screen $d\Sigma=dAdt$
\begin{eqnarray}\label{Q1}
\delta Q=\int_{\Sigma}J^aN_a dAdt,
\end{eqnarray}
where $J^a$ is electric current of matter in the bulk $M$.
Assume the surface electric current on the screen be $j_i$, then the
charge change on the screen is
\begin{eqnarray}\label{Q2}
\delta Q&=&\int u^ij_i|^{t+dt}_{t}-\int m^ij_i\sqrt{h}dydt\nonumber\\
&=&-\int_{\partial\Sigma}M^ij_i\sqrt{h}dy^2=-\int_{\Sigma}D^ij_i dA dt,
\end{eqnarray}
where the first term in the first equality is due to change of charge
density and the second term is the electric flow through the patch
boundary parameterized by $y$. Again these two terms can be naturally written in a uniform form $-\int_{\partial\Sigma}M^ij_i\sqrt{h}dy^2$.
Applying Stokes's Theorem, we get the last equality.

According to the holographic principle, we should equate Eq.(\ref{Q1}), the charge flow through the patch of the holographic screen, and Eq.(\ref{Q2}),
the charge change on this patch, yielding
\begin{eqnarray}\label{flow}
D^ij_i=-N_aJ^a.
\end{eqnarray}
So far, we have not made any assumption about the form of $j_i$. Now, let us gain some insight into the form of $j_i$ from Eq.(\ref{flow}).
According to the holographic principle, we expect to derive the bulk equations of motion from Eq.(\ref{flow}), which implies that $j^i=e^a_ij^a$
should linearly depend on $N^a$. Thus, the general form for $j_a$ is
\begin{eqnarray}\label{j}
j_a=A_{ab}N^b+A_{abc}\nabla^bN^c+...
\end{eqnarray}
And $j_a$ must satisfy the following conditions
\begin{eqnarray}\label{cond}
N^aj_a=0, \ \ D^ij_i=\gamma^{ab}\nabla_a j_b=-N_aJ^a,
\end{eqnarray}
where $\gamma^{ac}$ is the projection operator defined as
$\gamma^{ac}=g^{ac}-N^aN^c$. For simplicity, we consider the
simplest assumption $j_a=A_{ab}N^b$. The conditions Eq.(\ref{cond})
become
\begin{eqnarray}\label{cond1}
A_{ab}N^aN^b=0, \ \ (\gamma^{ac}\nabla_aA_{cb})N^b+\gamma^c_aA_{cb}\nabla^aN^b=-N^bJ_b.
\end{eqnarray}
From the last equation of Eq.(\ref{cond1}), we have
\begin{eqnarray}\label{cond2}
(\gamma^{ac}\nabla_aA_{cb}+J_b)N^b=0, \ \gamma^c_aA_{cb}\nabla^aN^b=0
\end{eqnarray}
Since $\gamma^c_a\nabla^aN^b=-K^{cb}$ is symmetrical in $c$ and $b$,
we derive $\gamma^a_c\gamma^d_b A_{ad}=-\gamma^a_b\gamma^d_c A_{ad}$
from $\gamma^c_aA_{cb}\nabla^aN^b=0$. For we can change the
direction of $N^a$ arbitrarily with changing the screen, taking
$\gamma^a_c\gamma^d_b A_{ad}=-\gamma^a_b\gamma^d_c A_{ad}$ and
$A_{ab}N^aN^b=0$ into account, we find that $A_{ab}$ is
antisymmetric. Thus, the first equation of Eq.(\ref{cond2}) becomes
\begin{eqnarray}\label{}
(\gamma^{ac}\nabla_aA_{cb}+J_b)N^b&=&(\nabla_aA^a_{b}+J_b)N^b-(N^a\nabla_aA_{cb})N^cN^b\nonumber\\
&=&(\nabla_aA^a_{b}+J_b)N^b=0.
\end{eqnarray}
So for arbitrary $N^b$, we get
\begin{eqnarray}\label{}
\nabla_aA^{ab}=-J^b,
\end{eqnarray}
with $A^{ab}$ an antisymmetric tensor and the surface electric
current $j_a=A_{ab}N^b$ on the screen. Note that conservation of
charge $\nabla_aJ^a=0$ is satisfied automatically for antisymmetric
$A^{ab}$
\begin{eqnarray}\label{}
-\nabla_aJ^a=\nabla_a\nabla_bA^{ab}=\frac{1}{2}(\nabla_a\nabla_b-\nabla_b\nabla_a)A^{ab}=R_{ab}A^{ab}=0.
\end{eqnarray}

The above approach can be directly extended to the case of magnetic
charge. Assume the magnetic current in the bulk and on the screen
be $J^a_m=0$ and $j^a_m=B^{ab}N_b$, respectively. With the same procedure we arrive at
\begin{eqnarray}\label{magnetic1}
B^{ab}=-B^{ba}, \ \ \nabla_aB^{ab}=-J^b_m=0.
\end{eqnarray}
Since the magnetic current  $J^b_m=0$ in the bulk, it is expected that
$j^a_m=B^{ab}N_b$ is not an independent physical quantity, it should
either vanish or be related to the surface electric current $j^a$ on
the screen. In view of the electromagnetical duality, it is natural
to assume that the surface electric current and  magnetic current on the
screen are related with each other by the Hodge duality
\begin{eqnarray}\label{duality}
B^{ab}=\frac{1}{2}\epsilon^{abcd}A_{cd}.
\end{eqnarray}
Then, Eq.(\ref{magnetic1}) becomes
\begin{eqnarray}\label{magnetic}
\epsilon^{abcd}\nabla_bA_{cd}=0,
\end{eqnarray}
which is just the Bianchi identity. The general solution of the
above equation is $A_{cd}=\partial_dA_c-\partial_cA_d$ with $A_c$ an
arbitrary vector field. Rename $A_{cd}$ by $F_{dc}$, let us
summarize our results. The surface electric current and  magnetic
current on the screen are
\begin{eqnarray}\label{current}
j^a=F^{ba}N_b,\ \ \ j^a_m=\frac{1}{2}\epsilon^{bacd}F_{cd}N_b.
\end{eqnarray}
The equations of motion in the bulk $M$ are
\begin{eqnarray}\label{Maxwell}
\nabla_aF^{ab}=J^b, \ \ \nabla_{[a}F_{bc]}=0,
\end{eqnarray}
where $F_{ab}=\partial_aA_b-\partial_bA_a$ and $[ \ \ ]$ denotes
complete antisymmetrization. The above equations are just the
Maxwell equations. Applying the formulas
\begin{eqnarray}\label{formulas}
F_{ab}\  ^{*}F^{bc}=\frac{1}{4}F\  ^{*}F\delta^c_a,\ \
F_{ab}F^{bc}-\ ^\ast F_{ab}\ ^\ast F^{bc}=\frac{1}{2}F^2\delta^c_a,
\end{eqnarray}
we obtain the following interesting identities
\begin{eqnarray}\label{identities}
j j_M=-\frac{1}{4}F\  ^{*}F,\ \ j^2-j^{2}_M=-\frac{1}{2}F^2,
\end{eqnarray}
where $^{*}F^{ab}$ is the Hodge duality of $F_{cd}$,
$^{*}F^{ab}=\frac{1}{2}\epsilon^{abcd}F_{cd}$. Note that the L.H.S
of Eqs.(\ref{identities}) are physical quantities on the screen while
the R.H.S of Eqs.(\ref{identities}) contain only physical quantities
in the bulk which are independent of the direction of the screen
$N^a$, so $j^a$ and $j^a_M$ contain all the information of the bulk
($F^2$ and $^*FF$) which is a reflection of the holographic
principle.

 Now, we have derived the Maxwell equations from the holographic
principle and an appropriate assumption for the relationship between
the surface electric current and magnetic current on the holographic
screen.

\section{Conclusions}

 We have derived f(R) gravity and the Maxwell
 equations from the holographic program we proposed in \cite{Li}. We find the surface
 stress tensor and surface electric current, surface magnetic current
 for f(R) gravity and Maxwell's theory, respectively. It is
 interesting to extend our holographic approach to more general higher derivative
 gravity, and investigate the corresponding thermodynamics on a
 time-like holographic screen. It should be mentioned that in Sect.4 we
 only find the simplest solution of Eqs.(\ref{j}) and
 (\ref{cond}), whether there are other solutions and corresponding holographic
 electromagnetic theories is an interesting problem. We hope we
 will gain more insight into these problems in the future.

\section*{Acknowledgements}
This research was supported by a NSFC grant No.10535060/A050207, a
NSFC grant No.10975172, a NSFC group grant No.10821504 and Ministry
of Science and Technology 973 program under grant No.2007CB815401.

\section*{Appendix}

 In this appendix, we shall prove that the second term $\xi^{a}\nabla_a(n K+q)$ in
Eq.(\ref{E2}) does not contain the term $\xi^{a}N^{b}B_{ab}$, where
$B_{ab}$ is independent of $\xi^{a}$ and $N^{b}$. If
$\xi^{a}\nabla_a(n K+g)$ does include such terms, $(n K+g)$ must
depend on $N^b$ linearly
\begin{equation}\label{}
n K+g=A_bN^b+A_{bc}\nabla^bN^{c}+...
\end{equation}
Thus,
\begin{eqnarray}\label{T0}
\xi^{a}\nabla_a(n
K+g)&=&\xi^aN^b\nabla_aA_b+\xi^aA^b\nabla_aN_b+(\xi^d\nabla_dA^{ab})\nabla_aN_b...\nonumber
\\
&=&\xi^aN^b\nabla_aA_b+\xi^d\nabla^aN^b(\nabla_dA_{ab}+g_{da}A_b)+...
\end{eqnarray}
Take into consideration the fact that Eq.(\ref{E1}) does not contain the term $\xi^d\nabla^aN^b$ and that $R_{ab}$ is
symmetrical, equating Eq.(\ref{E1}) and Eq.(\ref{E2}) yields
\begin{eqnarray}\label{}
\nabla_aA_b=\nabla_bA_a,
\end{eqnarray}
\begin{eqnarray}\label{}
D_iA_{jk}+\gamma_{ij}A_k=e^d_ie^a_je^b_k(\nabla_dA_{ab}+g_{da}A_b)=0.
\end{eqnarray}
From the above equations, we derive
\begin{eqnarray}\label{T1}
D_iA+A_i=0, \ \ \ D_iA^i_{k}+4A_k=0, \ \ A=A_{jk}\gamma^{jk},
\end{eqnarray}
\begin{eqnarray}\label{T2}
D_iA_{jk}-D_jA_{ik}=0,\ \ \ D_iA-D_jA^j_{i}=D_iA+4A_i=0.
\end{eqnarray}
From the first equation in Eq.(\ref{T1}) and the last equation in
Eq.(\ref{T2}), we get $A_i=0$, so
$A^a=e^a_iA^i+N^a(N_bA^b)=N^a(N_bA^b)$. Now, we can prove that the
first term of Eq.(\ref{T0}) vanishes:
\begin{eqnarray}\label{}
\xi^aN^b\nabla_aA_b&=&\xi^aN^b\nabla_bA_a=N^b\nabla_b(\xi_aA^a)-N^bA^a\nabla_b\xi_a\nonumber
\\
&=&N^b\nabla_b(\xi_iA^i)-(N_cA^c)N^bN^a\nabla_b\xi_a=0.
\end{eqnarray}

Thus, the second term $\xi^{a}\nabla_a(n K+g)$ in Eq.(\ref{E2}) does
not contain the term as $\xi^{a}N^{b}B_{ab}$. We must require
$(nK+g)=0$ in order to equate Eq.(\ref{E1}) and Eq.(\ref{E2}).





\bibliographystyle{elsarticle-num}
\bibliography{<your-bib-database>}



\end{document}